\documentclass[fleqn,usenatbib]{mnras}

\usepackage{newtxtext,newtxmath}

\usepackage[T1]{fontenc}
\usepackage{ae,aecompl}

\usepackage{graphicx}	
\usepackage{amsmath}	
\usepackage{amssymb}	
\usepackage{bm}
\usepackage{color}

\title[Polarized Balmer Line from SNR Shocks]{Polarized Balmer Line Emission from Supernova Remnant
Shock Waves Efficiently Accelerating Cosmic Rays}

\author[J. Shimoda et al.]{
Jiro Shimoda,$^{1}$\thanks{E-mail: s-jiro@phys.aoyama.ac.jp (JS)}
Yutaka Ohira,$^{1}$
Ryo Yamazaki,$^{1}$
J. Martin Laming$^{2}$
\newauthor
and Satoru Katsuda$^{3}$
\\
$^{1}$Depertment of Physics and Mathematics, Aoyama-Gakuin University, Sagamihara, Kanagawa 252-5258, Japan\\
$^{2}$Space Science Division, Naval Research Laboratory, Code 7684, Washington DC 20375, USA\\
$^{3}$Department of Physics, Faculty of Science \& Engineering, Chuo University, 1-13-27 Kasuga, Bunkyo, Tokyo 112-8551, Japan
}

\date{Accepted XXX. Received YYY; in original form ZZZ}

\pubyear{2017}

\begin{document}
\label{firstpage}
\pagerange{\pageref{firstpage}--\pageref{lastpage}}
\maketitle

\begin{abstract}
Linearly polarized Balmer line emissions from supernova remnant shocks are
studied taking into account the energy loss of the shock owing to the
production of nonthermal particles. The polarization degree depends on the
downstream temperature and the velocity difference between upstream and
downstream regions. The former is derived once the line width of the broad
component of the H~$\alpha$ emission is observed. Then, the observation of
the polarization degree tells us
the latter.
At the same time, the estimated value of the velocity difference independently
predicts
adiabatic downstream temperature that is derived from Rankine-Hugoniot relations for adiabatic shocks.
If the actually observed downstream temperature is lower than
the adiabatic temperature,
there is a missing thermal energy which is consumed for particle
acceleration. It is shown that a larger energy loss rate leads to more
highly polarized H~$\alpha$ emission. Furthermore, we find that polarized
intensity ratio of H~$\beta$ to H~$\alpha$ also depends on the energy loss
rate and that it is independent of uncertain quantities such as electron
temperature, the effect of Lyman line trapping and our line of sight.
\end{abstract}




\section{Introduction}
\label{sec:intro} Supernova remnants (SNRs) are the best candidate sites for
Galactic cosmic-ray (CR) production. Measurements of Galactic CR energy
density around the Earth require that roughly a tenth of supernova explosion
energy is consumed for CR acceleration. The CR acceleration efficiency at
SNRs is estimated as an energy loss rate of the SNR shock wave
\citep[e.g.][]{hughes00,warren05,tatischeff07,helder09,helder13,morlino13a,morlino14}.
Since the shock loses its kinetic energy due to the CR acceleration, the
downstream temperature becomes lower than that in the adiabatic case.
Therefore, if we measure both the downstream temperature ($T_{\rm down}$) and
the shock velocity ($V_{\rm sh}$) independently, we can estimate the energy
loss rate. In order to do this, we define $kT_{\rm RH}=\frac{3}{16}\mu m_{\rm p}V_{\rm sh}{}^2$
($k$ is Boltzmann constant and $\mu$ is the mean molecular weight),
which is the adiabatic downstream temperature
predicted by Rankine-Hugoniot relations in the strong shock limit without CR
acceleration (that is, the adiabatic shocks). Then, the energy loss rate is
defined as \citet{shimoda15}
%
\begin{eqnarray}
\eta=\frac{T_{\rm RH}-T_{\rm down}}{T_{\rm RH}}.
\label{eta}
\end{eqnarray}
%
Observations of SNR~RCW~86 give an example. The shock velocity is measured
by the proper motion of an H~$\alpha$ filament as $V_{\rm sh}\approx1800~{\rm
km/s}$, that gives $T_{\rm RH}\approx4~{\rm keV}(\mu/0.62)(V_{\rm sh}/1800~{\rm km~s^{-1}})^2$
\citep{helder13}.
On the other hand, the downstream temperature is derived from spectroscopy of the
H~$\alpha$ emission as $T_{\rm down}\approx2~{\rm keV}$ \citep{helder09}.
Combining these observations, we obtain $\eta\approx0.5$.
\footnote{
In \citet{helder13}, the proper motions of H~$\alpha$ filaments were observed as
$1871\pm250,~1196\pm367$ and $1325\pm221~{\rm km~s^{-1}}$ at the region where the downstream temperature was measured.
The adiabatic downstream temperatures are calculated as $kT_{\rm RH}=4.5\pm1.2,~1.8\pm1.1$ and $2.3\pm0.7~{\rm keV}$, respectively,
resulting in $\eta=0.6\pm0.1$, $-0.08\pm0.7$ and $0.1\pm0.3$, respectively.}
Such extremely high
energy loss rate would alter the long-term evolution of the shock
\citep[e.g.][]{cohen98,liang00}. However, to evaluate the shock velocity from
the proper motion measurements, we need a distance to the SNR with high
accuracy, which is often hard in astronomy. In this paper, following work by
\citet{laming90}, we show that the energy loss rate can be obtained by
polarization degree of the H~$\alpha$ emissions without precise measurements
of the distance.
\par
We will briefly review the H~$\alpha$ emissions in SNRs. The young SNR shock
is formed by the interaction between charged particles and plasma waves
rather than particle Coulomb collision processes (so called collisionless
shock). The shock wave propagates into the interstellar medium (ISM), which
is in general partially ionized. The charged particles in ISM are heated by
the collisionless shock wave, while the neutral particles (hereafter, we
consider only hydrogen atoms) are not affected. Therefore, the hydrogen atoms
collide with charged particles in the downstream region owing to a finite
relative velocity. As a result, the hydrogen atoms entering the downstream
region are excited, radiating Balmer line emissions, and they are eventually
ionized. Since the length of the emitting region, which is on the order of
the mean free path of atomic collision, $\sim10^{16}~{\rm cm}$, is much
shorter than the radius of SNR ($\sim1\mathchar`-10~{\rm pc}$), the Balmer
line emissions are bright along the shock surface on the sky. Such SNR shocks
are called as Balmer dominated shocks \citep[BDSs,
e.g.][]{chevalier78,chevalier80,laming90,heng07,heng10,adelsberg08,morlino12}.
In addition, the H~$\alpha$ emissions from the upstream region have been observed in some SNRs
\citep[e.g.][]{ghavamian00,lee07,lee10,katsuda16}.
The spectrum of the Balmer line emissions often consists of narrow and broad
components. The former is caused by the direct excitation via collision
between the hydrogen atoms and the charged particles. On the other hand, the
latter is emitted after the charge exchange reaction between the
hydrogen atoms entering the shock and the downstream heated protons. Hence,
the width of the narrow component reflects thermal/nonthermal velocity of
upstream hydrogen atoms \citep[often observed as $\sim20\mathchar`-50~{\rm
km~s^{-1}}$:][]{medina14,knezevic16}, and the width of the broad component
reflects the downstream proton temperature \citep[e.g. $\sim2000~{\rm
km~s^{-1}}$:][]{chevalier80}. Thus, the downstream proton temperature can be
directly measured from the width of broad H~$\alpha$ emission
\citep{chevalier80}.
It is also possible to measure the electron temperature,
the heating process of electrons in the formation of collisionless shock is
a matter of debate and still widely studied
\citep[e.g.][]{cargill88,ghavamian01,ghavamian02,ohira07,ohira08,rakowski08}.
In BDS, the intensity ratio of the broad to the narrow component depends on
temperature equilibration between ions and electrons \citep[e.g.][]{adelsberg08}.
Therefore, the electron temperature is
derived by the intensity ratio of the broad to the narrow component
\citep[e.g.][]{ghavamian01,ghavamian02,adelsberg08,morlino12,morlino13b}. BDS
is seen in a number of SNRs. Moreover, measurements of the Balmer line nature
could be an essential probe of the collisionless shock physics.
\par
Some of the of hot hydrogen atoms emerging from the charge-exchange reaction
can leak upstream because they do not feel electromagnetic fields.
The leaking hot hydrogen atoms are ionized again through the collision with incoming
charged particles. As a result, the hot protons are injected in the upstream
region \citep[][]{lim96,blasi12,ohira12}. The injected protons can be
scattered by electromagnetic waves in the region adjacent to the shock and
accelerated by diffusive shock acceleration
\citep[e.g.][]{ohira12,ohira13,ohira16a,ohira16b}. Therefore, BDSs are
expected to be an accelerator of CR protons.
\par
Recently, \citet{sparks15} discovered linearly polarized H~$\alpha$ emission
with $2.0\pm0.4$ per cent polarization degree in the north-west region of the young
SNR, SN~1006, in good agreement with the original prediction of
\citet{laming90}. In laboratory experiments, linearly polarized H~$\alpha$
emission from hydrogen atoms in electron beams has been measured with
$\sim40$ per cent polarization degree \citep[e.g.][]{kleinpoppen68}. The
measurements of polarized atomic lines act as strong tools to study atomic
structure. The electron beam, which collides with hydrogen atoms from only
one direction, behaves as a quantization axis of the orbital angular momentum
of bound electron in excited hydrogen atoms \citep[e.g.][]{takacs96}.
The collisional excitation is essentially nonrelativistic and can be
discussed in terms of orbital angular momenta.
Once excited the orbital angular momentum of the electron couples
to its spin angular momentum to form a total angular momentum, $j$, with $z$-component $m_j$.
The bound electrons lose their energy and total angular momentum owing to the
spontaneous transition, and emit photons. The polarization of the photon is
then determined to be linear or circular by a variation of the orbital
angular-momentum component along with the beam direction, which is given by
magnetic quantum number $m_j$. The linearly polarized intensity becomes
largest when viewed from the direction orthogonal to the beam. For BDS,
charged hot particles hit cold hydrogen atoms from various directions in
the downstream region. In the rest frame of hydrogen atoms (i.e. the upstream
frame), the colliding charged particles are seen as a mildly-collimated beam.
Therefore, this anisotropy eventually causes the net polarization of the line
emission, with the polarization degree of a few percent.
\par
If the SNR shock with shock velocity $V_{\rm sh}$ efficiently accelerates CRs, then they can escape the shock,
carrying away significant energy. As a result, the downstream temperature
becomes lower than $T_{\rm RH}$, yielding larger anisotropy of the particle velocity
downstream. \citet{laming90} studied the linearly polarized H~$\alpha$
emission from BDS without CR acceleration. Then, he showed that a
few$\mathchar`-10$ per cent polarization degree can be observed. His study was
limited owing to the lack of atomic data on proton collisional excitation
cross section and the line of sight direction was fixed as orthogonal to the
shock normal, which gives the largest linear polarization degree.
\citet{heng08} and \citet{tseli12} updated the atomic data and fitting
functions. Using their data, we study the linearly polarized Balmer line
emissions from the SNR shocks losing their thermal energy. We show that a
higher energy loss rate causes higher polarization degree. The polarization
degree of the line emission is determined by the anisotropy of the velocity
distribution of charged particles (i.e. collimation of incident beam), which
is given by the downstream temperature and the upstream fluid velocity. We
can measure the downstream temperature from the width of the broad H~$\alpha$
line, whereas the downstream fluid velocity is derived from the polarization
measurements. Since the shock velocity $V_{\rm sh}$ relates $T_{\rm RH}$ by
Rankine-Hugoniot relation, we can obtain the energy loss rate $\eta$ without
measuring of SNR distance. In Sect. 2, we formulate the polarized Balmer line
emissions from the shock accelerating nonthermal particles. In Sect. 3, we
present the results of polarization degree of H~$\alpha$. In Sect. 4, the
polarized intensity ratio of H~$\beta$ to H~$\alpha$ is discussed. Finally,
we summarize our results and discuss on future prospects for the estimation
of $\eta$.

\section{Physical Model}
\label{sec:physical model} In \citet{laming90}, only the case of viewing
angle orthogonal to the shock normal was considered. In this paper, we extend
his study, and investigate the linearly polarized Balmer-emission from the
shock, which loses kinetic energy due to CR acceleration, with arbitrary
viewing angle. In the following, we consider only the narrow component of
H~$\alpha$ and H~$\beta$ emissions resulting from the direct collisional
excitation and denoting by ``n''.

\subsection{The Model Geometry}
\begin{figure}
	\includegraphics[width=\columnwidth]{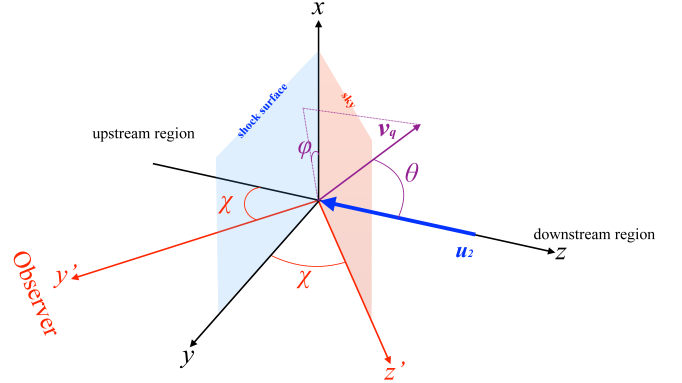}
    \vspace{4cm}
    \caption{Schematic diagram of the shock geometry.
    The blue $x\mathchar`-y$ plane ($z=0$) represents the shock surface.
    The blue arrow $\bm{u_2}$ shows the downstream velocity
    in the upstream rest frame, which is parallel to the $z$ axis.
    The purple vector $\bm{v_q}$ is the velocity of the particle $q$,
    that collides with the hydrogen atom at the origin.
    The red $y'$ axis is parallel to the line of sight and the red $z'$ axis is orthogonal
    to $x$--$y'$ plane, and makes an angle $\chi$ with the $z$ axis.
    The red $x\mathchar`-z'$ plane represents the plane of the sky, which is
    orthogonal to the line of sight.}
    \label{fig:coordinate}
\end{figure}

Figure \ref{fig:coordinate} shows the schematic diagram of the shock
geometry. The blue sheet is $z=0$ plane and represents the shock surface. The
blue arrow shows the downstream velocity in the upstream rest frame, which is
parallel to the $z$ axis. The purple vector is the velocity of the particle $q$
that collides with the hydrogen atom at the origin,
%
\begin{eqnarray}
\bm{v_q}=(v_q\sin\theta\cos\varphi,v_q\sin\theta\sin\varphi,v_q\cos\theta).
\nonumber
\end{eqnarray}
The red $y'$ axis is parallel to the line of sight and the red $z'$ axis is
perpendicular, which makes an angle $\chi$ to the $y$ axis. The red sheet
represents the plane of the sky, which is orthogonal to the line of sight.
%
%
\subsection{The Polarized Line Emission}
Polarized atomic line emission induced by collisional excitation is reviewed
by \citet{percival58}. In this paper, we treat the dipole transition that
makes Balmer line emission.
\par
Let $\sigma_{nlm_l,q}$ be the cross section of collisional excitation by
the particle $q$ from the ground state hydrogen atom to the excited state
$nlm_l$, where $n$ is the principal quantum number, $l=0,1,...,n-1$ is the
orbital angular momentum quantum number and $m_l=-l,-l+1,...,l$ is the
magnetic quantum number. We evaluate the orbital angular momentum of the
bound electron of the hydrogen atom along the incident direction of the
particle $q$. Then, the quantum number $l$ represents the orbital angular
momentum magnitude of the bound electron, $L=\sqrt{l(l+1)}\hbar$, while the
magnetic quantum number $m_l$ gives the component of the orbital angular
momentum parallel to $\bm{v_q}$, $L_r=m_l\hbar$. Let $A_{njm_j,n'j'm_j'}$ be
the spontaneous transition rate per unit time from the atomic state of
$njm_j$ to $n'j'm_j'$. The total angular momentum $j$ is formed by vector
addition of $l$ and $s$, the orbital and spin angular momenta, and this
coupling introduces some depolarization. The spontaneous transition is only
allowed for $\Delta l=l-l'=\pm1$ and $|\Delta m_l|=|m-m'|\le1$, $\Delta
j\le1$ and $|\Delta m_j|\le1$. In what follows however, we adopt a
nonrelativistic description of the hydrogen atom for decay rates and
branching ratios, i.e. using $A_{nlm_l,n'l'm_l'}$ instead of
$A_{njm_j,n'j'm_j'}$, and we drop the $l$ or $j$ subscript on $\Delta m$.
\par
By the conservation of the angular momentum, the polarization of emitted
photon is characterized by the subtraction of the eigenvalues of $L_r$ before
and after the transition. We presume that the transition of the hydrogen atom
induces second time derivative of electric dipole moment whose polarization
vector is $\hat{\ddot{d}}_{\Delta m}$. The polarization vector can be written
as
%
\begin{eqnarray}
\hat{\ddot{d}}_{0} &=& \hat{v}_{q,r}e^{{\rm i}\omega_{\rm B} t}, \nonumber \\
\hat{\ddot{d}}_{\pm1} &=& \frac{1}{\sqrt{2}}(\hat{v}_{q,\theta}\pm {\rm i}\hat{v}_{q,\varphi})e^{{\rm i}\omega_{\rm B} t},
\end{eqnarray}
%
where unit vectors are defined as
%
\begin{eqnarray}
\hat{v}_{q,r} &=& (\sin\theta\cos\varphi,\sin\theta\sin\varphi,\cos\theta), \nonumber \\
\hat{v}_{q,\theta} &=& (\cos\theta\cos\varphi,\cos\theta\sin\varphi,-\sin\theta), \nonumber \\
\hat{v}_{q,\varphi} &=&
\hat{v}_{q,r}\times\hat{v}_{q,\theta}= (-\sin\varphi,\cos\varphi,0), \nonumber
\end{eqnarray}
%
and ${\rm i}$ is imaginary unit, and $\omega_{\rm B}$ is the angular frequency of
the Balmer-series emission. The electric field of the photon emitted along
the line of sight direction is given by
%
\begin{eqnarray}
\bm{E}_{\Delta m}(t)=\left\{\hat{y'}\times(\hat{y'}\times\hat{\ddot{d}}_{\Delta m})\right\}E(t),
\end{eqnarray}
%
where the unit vector along the line of sight is
%
\begin{eqnarray}
\hat{y'}=(0,\sin\chi,-\cos\chi),
\nonumber
\end{eqnarray}
%
and $E(t)$ is the electric field strength.
We decompose the observed electric field as
%
\begin{eqnarray}
\bm{E}_{\Delta m,z'} &=& (\bm{E}_{\Delta m}\cdot\hat{z'})\hat{z'},
\nonumber \\
\bm{E}_{\Delta m,x } &=& (\bm{E}_{\Delta m}\cdot\hat{x })\hat{x},
\nonumber
\nonumber
\end{eqnarray}
%
where the basic vectors are written as
%
\begin{eqnarray}
\hat{z'} &=& (0,\cos\chi,\sin\chi),
\nonumber \\
\hat{x} &=& (1,0,0).
\nonumber
\end{eqnarray}
%
\par
The observed intensity of the line emission is proportional to the number of
hydrogen atoms that yield $\hat{\ddot{d}}_{\Delta m}$. Let $\sigma'_{\Delta
m,q}$ be the cross section inducing $\hat{\ddot{d}}_{\Delta m}$ resulting
from the collision between the particle $q$ and the hydrogen atom as
%
\begin{eqnarray}
\sigma'_{\Delta m,q}(v_q) &=& \sum_{\substack{l'=l\pm1 \\ m_l'=m_l+\Delta m}}
B_{nlm_l,n'l'm_l'}\sigma_{nlm_l,q}(v_q), \nonumber \\
B_{nlm_l,n'l'm_l'} &=& \frac{A_{nlm_l,nl'm_l'}}{\displaystyle\sum_{n',l'm_l'}A_{nlm_l,n'l'm_l'}},
\end{eqnarray}
%
where $B_{nlm_l,n'l'm_l'}$ is the branching ratio of the spontaneous
transition from the atomic level $nlm_l$ to $n'l'm_l'$. For fixed $n$ and
$n'$, we take the summation of $\sigma_{\Delta m,q}$ for $l,~l',~m,~m'$ under
the constraints $l-l'=\pm1$, $\Delta m=0~{\rm or}~\pm1$. In the following, we
regard $\sigma'_{1,q}$ as identical to $\sigma'_{-1,q}$ because the
collision between the particle $q$ and the hydrogen atom is axially
symmetric. The Stokes parameters of the observed line emission are written as
%
\begin{eqnarray}
& Q_{\rm n} = \langle E_{{\rm obs},z'}E_{{\rm obs},z'}{}^*\rangle
-\langle E_{{\rm obs},x}E_{{\rm obs},x}{}^*\rangle, \nonumber \\
& I_{\rm n} = \langle E_{{\rm obs},z'}E_{{\rm obs},z'}{}^*\rangle
+\langle E_{{\rm obs},x}E_{{\rm obs},x}{}^* \rangle, \nonumber
\end{eqnarray}
%
where $E_{{\rm obs},z'}$ ($E_{{\rm obs},x}$) is $z'$ ($x$) components
of the observed electric field,
the asterisk $^*$ represents the complex conjugate,
and $\langle E E^*\rangle=\int_0^T EE^*/Tdt$ means long-time average
in the random phase approximation.
Let $f_q(\bm{v_q},\bm{u_2})$ be a velocity distribution function of particle $q$.
We approximate the velocity distribution function of hydrogen atom as Dirac delta function, $\delta(\bm{v_{\rm H}})$.
Then, the observed Stokes parameters are
%
\begin{eqnarray}
\begin{split}
& Q_{\rm n}
= n_{\rm H}\sum_{q} n_q \int v_q f_q(\bm{v_q},\bm{u_2}) \\
& \times \left[
\sigma'_{0,q}|\bm{E}_{0,z'}|^2 +
\sigma'_{1,q}|\bm{E}_{1,z'}|^2 +
\sigma'_{-1,q}|\bm{E}_{-1,z'}|^2
\right. \\
& \left. -
\left\{
\sigma'_{0,q}|\bm{E}_{0,x}|^2 +
\sigma'_{1,q}|\bm{E}_{1,x}|^2 +
\sigma'_{-1,q}|\bm{E}_{-1,x}|^2
\right\}
\right]d^3\bm{v_q} \\
& = n_{\rm H}E^2 \sum_{q}n_q\int v_q f_q(\bm{v_q},\bm{u_2}) \\
& \times \left[
\sigma'_{0,q}|\hat{z'}\cdot\hat{v}_{q,r}|^2 +
\sigma'_{1,q}
\left(
|\hat{z'}\cdot\hat{v}_{q,\theta}|^2+|\hat{z'}\cdot\hat{v}_{q,\varphi}|^2
\right) \right. \\
& \left. -
\left\{
\sigma'_{0,q}|\hat{x}\cdot\hat{v}_{q,r}|^2 +
\sigma'_{1,q}
\left(
|\hat{x}\cdot\hat{v}_{q,\theta}|^2+|\hat{x}\cdot\hat{v}_{q,\varphi}|^2
\right)
\right\}
\right]d^3\bm{v_q},
\end{split}
\label{Q_s}
\end{eqnarray}
%
and likewise
%
\begin{eqnarray}
\begin{split}
& I_{\rm n}
= n_{\rm H}E^2 \sum_{q} n_q \int v_q f_q(\bm{v_q},\bm{u_2}) \\
& \times \left[
\sigma'_{0,q}|\hat{z'}\cdot\hat{v}_{q,r}|^2 +
\sigma'_{1,q}
\left(
|\hat{z'}\cdot\hat{v}_{q,\theta}|^2+|\hat{z'}\cdot\hat{v}_{q,\varphi}|^2
\right) \right. \\
& \left. +
\left\{
\sigma'_{0,q}|\hat{x}\cdot\hat{v}_{q,r}|^2 +
\sigma'_{1,q}
\left(
|\hat{x}\cdot\hat{v}_{q,\theta}|^2+|\hat{x}\cdot\hat{v}_{q,\varphi}|^2
\right)
\right\}
\right]d^3\bm{v_q},
\end{split}
\label{I_s}
\end{eqnarray}
%
where $n_{\rm H}$ and $n_q$ are the number density of the hydrogen atom and
particle $q$.  In the following, we consider only protons (denoted as ``p'')
and electrons (denoted as ``e'') as the particle $q$ which excite the hydrogen
atoms (i.e. $q=\{{\rm p,e}\}$ and $n_{\rm p}=n_{\rm e}$). When the ionization
degree of the upstream medium is significantly low, the collisional
excitation by hot hydrogen atoms emerged from the charge-exchange reaction
would also contribute to the production of Balmer photons. Indeed, the cross
section of the collisional excitation on the impact between proton and
hydrogen atom is comparable with that between two hydrogen atoms
\citep[e.g.][]{barnett90}. We neglect this process for simplicity.

\subsection{Cross Sections of Impact Excitation}
\label{sec:crosec} In order to calculate the Stokes parameters from Eqs.
\eqref{Q_s} and \eqref{I_s}, the data for the cross section, $\sigma_{\Delta
m_l,q}'$, are required. In the laboratory experiment, the values of
$\sigma_{\Delta m_l,q}'$ are derived by measuring the polarization degree of
Balmer emissions resulting from the collision between hydrogen atoms and a
charged particle beam \citep[e.g.][]{McConkey88}.
\par
%
\begin{figure}
    \center
	\includegraphics[width=10cm]{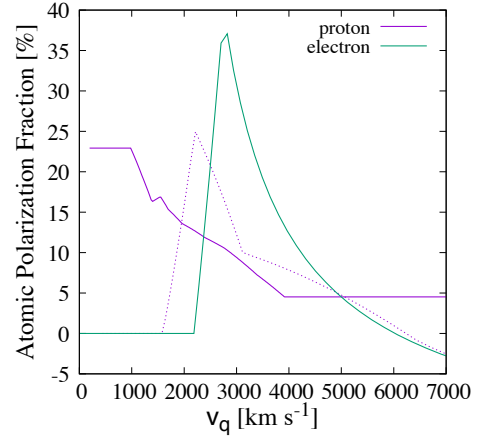}
    \caption{The atomic polarization fraction as a function of the colliding particle velocity $v_q$
    for the proton (magenta line) and electron impacts (green line).
    The dotted line was assumed in \citet{laming90} for proton impact.}
    \label{fig:apf}
\end{figure}
%
If we set $\chi=\pi/2$ and $f_q=\delta(\bm{v_q}-\bm{u_2})$
with $\bm{u_2}=(0,0,u_2)$ in the Eqs. \eqref{Q_s} and \eqref{I_s},
then the observed polarization degree is derived as
%
\begin{eqnarray}
\frac{Q_{\rm n}}{I_{\rm n}}=\sum_q\frac{\sigma'_{0,q}-\sigma'_{1,q}}{\sigma'_{0,q}+\sigma'_{1,q}}.
\nonumber
\end{eqnarray}
%
Thus, the polarization degree, $P_q$, of the line emission radiated from the hydrogen atom
in the direction perpendicular to the incident direction of particle $q$
is written as
%
\begin{eqnarray}
P_q=\frac{\sigma'_{0,q}-\sigma'_{1,q}}{\sigma'_{0,q}+\sigma'_{1,q}},
\end{eqnarray}
%
which is called the atomic polarization fraction.
In the following, we use the notations
$({\rm s,p,d,f,...})=(0,1,2,3,...)$ as the orbital angular momentum quantum number,
$l$, which are often used in atomic spectroscopy and astronomy.
\citet{percival58} and \citet{syms75} theoretically gave the fraction of H~$\alpha$ as
%
\begin{eqnarray}
\nonumber
& P_q({\rm H_\alpha})=
\left[
B_{3{\rm p},2{\rm s}}\frac{\sigma_{3{\rm p}0,q}-\sigma_{3{\rm p}\pm1,q}}{2}+
57\frac{\sigma_{3{\rm d}0,q}+\sigma_{3{\rm d}\pm1,q}-2\sigma_{3{\rm d}\pm2,q}}{100}\right] \\
\nonumber
& \times
\left[
\sigma_{3{\rm s}0,q}+
B_{3{\rm p},2{\rm s}}\frac{7\sigma_{3{\rm p}0,q}+11\sigma_{3{\rm p}\pm1,q}}{6} \right. \\
& \left.
+\frac{119\sigma_{3{\rm d}0,q}+219\sigma_{3{\rm d}\pm1,q}+162\sigma_{3{\rm d}\pm2,q}}{100}
\right]^{-1},
\end{eqnarray}
%
where $\sigma_{nl\pm m,q}=\sigma_{nl+m,q}+\sigma_{nl-m,q}$.
The numerical coefficients are considering the spin-orbit interaction,
but neglecting hyperfine structure. 
For the proton impact in the range of $1000~{\rm km~s^{-1}}\la v_{\rm p}\la4000{\rm~km~s^{-1}}$,
we can use the data derived by \citet{tseli12}. \citet{balanca98} showed that the atomic polarization
fraction of H~$\alpha$ is almost constant ($\approx0.25$) for $v_{\rm
p}\la1000{\rm~km~s^{-1}}$. Thus, we assume $P_{\rm p}({\rm
H_\alpha})\big|_{v_{\rm p}\le1000{\rm~km~s^{-1}}} =P_{\rm p}({\rm
H_\alpha})\big|_{v_{\rm p}=1000{\rm~km~s^{-1}}}$ and $P_{\rm p}({\rm
H_\alpha})\big|_{v_{\rm p}\ge4000{\rm~km~s^{-1}}} =P_{\rm p}({\rm
H_\alpha})\big|_{v_{\rm p}=4000{\rm~km~s^{-1}}}$. For the fraction from
electron impact, we follow the approximation by \citet{laming90} given as
%
\begin{eqnarray}
P_{\rm e}({\rm H_\alpha})=
\begin{cases}
0~~~{\rm for}~~~E_{\rm e}<0.5, \\
\frac{4-3\ln E_{\rm e}}{14.3+11\ln E_{\rm e}}
~~~{\rm for}~~~0.794\le E_{\rm e}, \\
1.36(E_{\rm e}-0.5)~~~{\rm for~~~0.5\le E_{\rm e}\le0.794}~~~,
\end{cases}
\label{P_e}
\end{eqnarray}
%
where $E_{\rm e}$ is the collision energy of the electron in the rest frame
of hydrogen atom (in atomic units). The atomic polarization fraction of
H~$\beta$ is hardly studied, compared with H~$\alpha$ and Ly~$\alpha$. On the
other hand, the fraction of Ly~$\beta$ is almost the same as Ly~$\alpha$
\citep{balanca98}. In the following, we assume the polarization fraction of H~$\beta$ is
the same as that of  H~$\alpha$. Figure \ref{fig:apf} shows the atomic
polarization fraction following proton (magenta line) and the electron
impacts (green line). The dotted line represents the fraction for the proton
impact assumed in \citet{laming90}. With the updated data of the atomic
polarization fraction, we obtain smaller polarization degree compared with
the previous work by \citet{laming90} at high proton temperatures, and
larger polarization at low proton temperature where he assumed the
polarization to be zero. Since the total cross section yielding the line
emission on particle $q$ impact is $\sigma_{{\rm
tot},q}=\sigma'_{0,q}+2\sigma'_{1,q}$, we can derive \citep{laming90}
%
\begin{eqnarray}
&& \sigma'_{0,q}+\sigma'_{1,q}
=\frac{2}{3-P_q}\sigma_{{\rm tot},q}, \\
&& \sigma'_{0,q}-\sigma'_{1,q}=P_q(\sigma'_{0,q}+\sigma'_{1,q}).
\end{eqnarray}
%
\par
%
\begin{table}
	\centering
	\caption{The branching ratio of H~$\alpha$ and H~$\beta$ \citep[e.g.][]{heng08}.}
	\label{tab:branch}
	\begin{tabular}{cc} 
        \hline
		\hline
		$B_{3{\rm p},2{\rm s}}$ & 0.1183 \\
		$B_{4{\rm s},2{\rm p}}$ & 0.5841 \\
        $B_{4{\rm s},3{\rm p}}$ & 0.4159 \\
        $B_{4{\rm p},1{\rm s}}$ & 0.8402 \\
		$B_{4{\rm p},2{\rm s}}$ & 0.1191 \\
        $B_{4{\rm p},3{\rm s}}$ & 3.643$\times10^{-2}$ \\
        $B_{4{\rm p},3{\rm d}}$ & 4.282$\times10^{-3}$ \\
		$B_{4{\rm d},2{\rm p}}$ & 0.7456 \\
        $B_{4{\rm d},3{\rm p}}$ & 0.2544 \\
		\hline
	\end{tabular}
\end{table}
%
The total cross section $\sigma_{{\rm tot},q}$ is the summation of
$B_{nl,n'l'}\sigma_{nl,q}^*$, where $B_{nl,n'l'}$ is the branching ratio of
the spontaneous transition from the atomic state $nl$ to $n'l'$, which is
summarized in Table \ref{tab:branch}. The $\sigma_{nl,q}^*$ is the effective
cross section for particle $q$ impact on the ground state hydrogen
including the effect of cascading from higher atomic levels.
Here, we omit the magnetic quantum number $m$ because it does not contribute the total cross
section and the branching ratio.
The total cross sections inducing the H~$\alpha$ emission are written as
%
\begin{eqnarray}
\sigma_{{\rm tot},q}\big|_{Q_{\rm n}({\rm H_\alpha})}
&=&\sigma_{3{\rm s},q}+B_{3{\rm p},2{\rm s}}\sigma_{3{\rm p},q}+\sigma_{3{\rm d},q}, \\
\sigma_{{\rm tot},q}\big|_{I_{\rm n}({\rm H_\alpha})}
&=&\sigma_{3{\rm s},q}^{*}+B_{3{\rm p},2{\rm s}}\sigma_{3{\rm p},q}^{*}+\sigma_{3{\rm d},q}^{*},
\label{sigma n=3 for I} \\
\sigma_{3{\rm s},q}^{*}
&=&\sigma_{3{\rm s},q}+B_{4{\rm p},3{\rm s}}\sigma_{4{\rm p},q},
\label{sigma 3s for I} \\
\sigma_{3{\rm p},q}^{*}
&=&\sigma_{3{\rm p},q}+B_{4{\rm s},3{\rm p}}\sigma_{4{\rm s},q}+B_{4{\rm d},3{\rm p}}\sigma_{4{\rm d},q},
\label{sigma 3p for I} \\
\sigma_{3{\rm d},q}^{*}
&=&\sigma_{3{\rm d},q}+B_{4{\rm p},3{\rm d}}\sigma_{4{\rm p},q}+\sigma_{4{\rm f},q},
\label{sigma 3d for I}
\end{eqnarray}
%
where we assume the emission resulting from the cascade from
the level with $n>3$ is unpolarized.
The cascade affects the observed polarization by a factor of $\sim5$ per cent
\citep{laming90}. Neglecting the cascades from higher atomic levels, we give
the total cross sections inducing the H~$\beta$ emissions as
%
\begin{eqnarray}
\sigma_{{\rm tot,}q}\big|_{I_{\rm n},Q_{\rm n}({\rm H_\beta})}
=B_{4{\rm s},2{\rm p}}\sigma_{4{\rm s},q}+B_{4{\rm p},2{\rm s}}\sigma_{4{\rm p},q}
+B_{4{\rm d},2{\rm p}}\sigma_{4{\rm d},q}.
\label{sigma n=4}
\end{eqnarray}
%

\par
%
\begin{figure}
    \center
	\includegraphics[width=10cm]{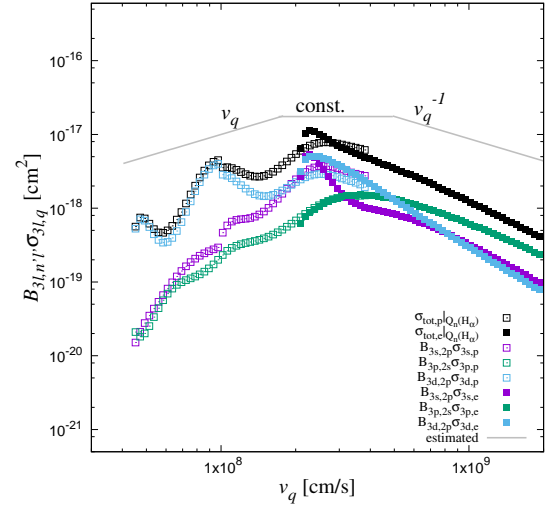}
    \vspace{10mm}
    \caption{The cross section of direct excitation to $n=3$ level for proton impact (open squares)
    and electron impact (closed squares), $B_{3l,n'l'}\sigma_{3l,q}$.
    The black squares are $\sigma_{{\rm tot},q}\big|_{Q_{\rm n}({\rm H_\alpha})}$.}
    \label{fig:cross3}
\end{figure}
%
%
\begin{figure}
    \center
    \includegraphics[width=10cm]{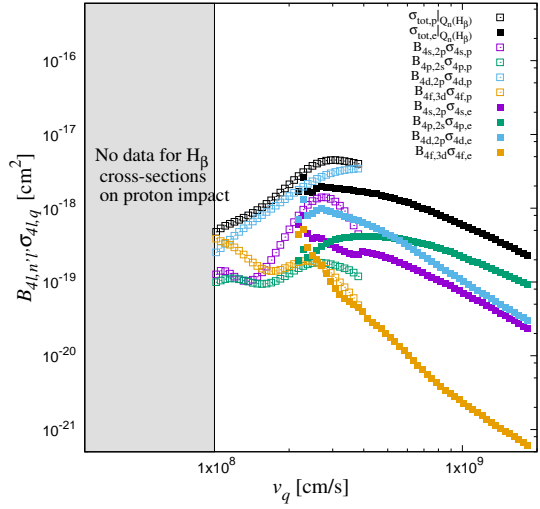}
    \vspace{10mm}
    \caption{The cross section for direct excitation to $n=4$ level by proton impact (open squares)
    and electron impact (closed squares), $B_{4l,n'l'}\sigma_{4l,q}$.
    The black squares are $\sigma_{{\rm tot},q}\big|_{Q_{\rm n}({\rm H_\beta})}$.}
    \label{fig:cross4}
\end{figure}
%
Figures \ref{fig:cross3} and \ref{fig:cross4} represent
$B_{nl,n'l'}\sigma_{nl,q}$ for H~$\alpha$ and H~$\beta$ emissions,
respectively. Here we take data from \citet{janev93}, \citet{bray95},
\citet{heng08} and \citet{tseli12}. For $v_{\rm p}\la1000{\rm~km~s^{-1}}$, we
use the data given by \citet{balanca98}, which were calculated with the
close-coupling approximation. This approximation is known to be applicable
for the range of $v_{\rm p}\ll\alpha c$ \citep[e.g.][]{tseli12}, where
$\alpha=1/137$ is the fine structure constant. In particular, the proton
cross section data in the range $1000{\rm~km~s^{-1}}\la v_{\rm
p}\la4000{\rm~km~s^{-1}}$ were derived by the direct numerical simulations by
\citet{tseli12}. Besides, we assume that the cross section for proton impact
excitation in the range $v_{\rm p}\ga4000~{\rm km~s^{-1}}$ is the same as
that of the electron impact excitation. Indeed, the proton impact cross
section for $n=3$ approaches to the electron's one for $v_{q}\ga3000~{\rm
km~s^{-1}}$ \citep[e.g.][]{janev93}. The fitting functions of these data are
provided by \citet{heng08} and \citet{tseli12}. The data for $n=4$ are
unavailable for the range $v_{\rm p}\la1000{\rm~km~s^{-1}}$. We treat the
cross section for proton impact to be zero in this range. The data for
electron impact and their fitting functions are provided by International
Atomic Energy Agency (https://www-amdis.iaea.org/ALADDIN/).
\par
The time scale of the spontaneous transition of the hydrogen atom from the
excited state to the ground state, $\sim10^{-8}\mathchar`-10^{-1}~{\rm s}$, is usually much
shorter than the mean collision time of particle $q$, $\sim10^{8}~{\rm s}
\left(\frac{n_q}{1~{\rm cm^{-3}}}\right)^{-1}
\left(\frac{\sigma}{10^{16}~{\rm cm^{2}}}\right)^{-1}
\left(\frac{v_q}{10^{8}~{\rm cm~s^{-1}}}\right)^{-1}$ for SNR shocks.
Therefore, we assume that all the hydrogen atoms are excited from the ground
state \citep[e.g.][]{adelsberg08}.

\subsection{Lyman Line Trapping}
\label{sec:trapping} A part of hydrogen atoms in the states $n>2$ emit
Lyman-series photons (e.g. $3{\rm p}\rightarrow1{\rm s}$). If the system is optically
thick for the Lyman photon, the emitted Lyman photons are absorbed by the
ground-state hydrogen atoms and eventually converted to other series as
Balmer, Paschen and so on \citep[e.g.][]{heng10}. In such a situation, for
instance, the branching ratio in Eq. \eqref{sigma n=3 for I} is effectively
$B_{3{\rm p},2{\rm s}}\approx1$ \citep[e.g.][]{adelsberg08}. It is called ``Case B''. On
the other hand, for optically thin limit (known as ``Case A''), we can use
the values of the branching ratio summarized in Table \ref{tab:branch}.
\par
In this paper, we assume that the Balmer photons emitted by the absorption of
Lyman photons are unpolarized.
Therefore, for Case B, the branching ratios concerning $I$
are approximately
%
\begin{eqnarray}
B_{3{\rm p},2{\rm s}} &=& 1, \nonumber \\
B_{4{\rm p},2{\rm s}} &=& 1-B_{4{\rm p},3{\rm s}}-B_{4{\rm p},3{\rm d}}, \nonumber \\
B_{4{\rm p},3{\rm s}} &=& 1-B_{4{\rm p},2{\rm s}}-B_{4{\rm p},3{\rm d}}. \nonumber
\end{eqnarray}
%

\subsection{Polarization from the Shock Wave}
Using the atomic data given in previous sections, we calculate the Stokes
parameters for an arbitrary velocity distribution of the particle $q$,
$f_q({\bm{v_q},\bm{u_2}})$. The velocity distribution function of particle
$q$ is set to a Maxwellian as
%
\begin{eqnarray}
f_q(\bm{v_q},\bm{u_2})=
\left(\frac{m_q}{2\pi kT_q}\right)^{\frac{3}{2}}
\exp\left(-\frac{m_q(\bm{v_q}-\bm{u_2})^2}{2kT_q}\right),
\label{f_s}
\end{eqnarray}
%
where $m_q$ and $k$ are respectively the mass of particle $q$ and Boltzmann constant,
$T_q$ is the downstream temperature of particle $q$.
Substituting Eq. \eqref{f_s} into Eqs. \eqref{Q_s}-\eqref{I_s}, and
integrating $0\le\theta\le\pi$ and $0\le\varphi\le2\pi$,
we derive
%
\begin{eqnarray}
\begin{split}
& Q_{\rm n}=4\pi n_{\rm H}E^2\sin^2\chi\sum_{q={\rm e,p}} n_q \left(\frac{D_q}{\pi}\right)^{\frac{3}{2}}
\frac{e^{-D_qu_2{^2}}}{(2D_qu_2)^4} \\
& \times \int_0^{\infty}\alpha_q{}^3
e^{-\left(\frac{D_q\alpha_q}{2u_2}\right)^2}
(\sigma_{0,q}-\sigma_{1,q}) \\
& \times \left[\left(\frac{3}{\alpha_q{}^3}+\frac{1}{\alpha_q}\right)\sinh\alpha_q
-\frac{3}{\alpha_q{}^3}\cosh\alpha_q
\right]d\alpha_q,
\end{split}
\label{Q}
\end{eqnarray}
%
and
%
\begin{eqnarray}
\begin{split}
& I_{\rm n}=4\pi n_{\rm H}E^2\sum_{q={\rm e,p}} n_q \left(\frac{D_q}{\pi}\right)^{\frac{3}{2}}
\frac{e^{-D_qu_2{^2}}}{(2D_qu_2)^4} \int_0^{\infty}\alpha_q{}^3
e^{-\left(\frac{D_q\alpha_q}{2u_2}\right)^2} \\
& \times
\left[(\sigma_{0,q}+\sigma_{1,q})
\frac{\sinh\alpha_q}{\alpha_q} \right.\\
& \left.
+ \left(\sigma_{0,q}-\sigma_{1,q}\right)
\times\left\{
\left(
\frac{1-3\cos^2\chi}{\alpha_q{}^3}-\frac{\cos^2\chi}{\alpha_q{}^2}
\right)\sinh\alpha_q \right. \right. \\
& \left. \left.
-\frac{1-3\cos^2\chi}{\alpha_q{}^2}\cosh\alpha_q
\right\}
\right]
d\alpha_q,
\end{split}
\label{I}
\end{eqnarray}
%
where
%
\begin{eqnarray}
D_q=\frac{m_q}{2kT_q}, \nonumber \\
\alpha_q=2 D_q u_2 v_q, \nonumber \\
u_2=|\bm{u_2}|. \nonumber
\end{eqnarray}
%
When $\chi=\pi/2$, Eqs. \eqref{Q} and \eqref{I} coincide with Eqs. (8) and
(9) of \citet{laming90}. In particular, when $\chi=\pi/2$ and $D_qu_2{}^2=0$,
we obtain $Q_{\rm n}=0$ and $I_{\rm
n}\propto\int_0^{\infty}(\sigma_{0,q}+2\sigma_{1,q})e^{-D_qv_q{}^2}v_q{}^3dv_q$,
that is, the observed emission is unpolarized due to the almost isotropic
collisions. On the other hand, when $D_qu_2{}^2\rightarrow\infty$ leading to
extremely anisotropic collisions, the observed emission is polarized as
$Q_{\rm n}/I_{\rm
n}=\sum(\sigma_{0,q}-\sigma_{1,q})/(\sigma_{0,q}+\sigma_{1,q})$.
\par
When we observe the shock from right in front (i.e. $\chi=0$),
the observed emission is unpolarized due to the isotropic collisions
between the particle $q$ and the hydrogen atom.

\subsection{Shock Jump Conditions}
To calculate the polarization degree from Eqs. \eqref{Q} and \eqref{I}, we
consider the downstream temperature $T_q$ and the downstream velocity $u_2$
measured in the upstream rest frame. Since the kinetic energy of the shock is
consumed for the acceleration of nonthermal particles, the downstream
temperature becomes lower than that in the adiabatic case, $T_{\rm RH}$. If
all the accelerated particles escape from the system, the shock dynamics can
be described like a radiative shock for optically thin limit.
\par
\citet{cohen98} analyzed the self-similar solution of the radiative shock.
Their analysis is independent of the details of the cooling process. They
considered that the cooling timescale is much shorter than the hydrodynamical
timescale and the shocked medium radiates a fixed fraction of its internal
energy in the cooling layer. In this case, the shock velocity is constant
during the time that a given fluid element crosses the radiative zone and
cools. Hence, the shock and the cooling layer are stationary. They
additionally assumed that the radiation does not affect the shock structure,
which remains adiabatic, and that the radiative layer follows it. In this
paper, we follow \citet{cohen98} to derive the shock jump conditions, and
assume that all the hydrogen atoms collide with the charged particles behind
the end of the cooling layer.
\par
Assuming a polytropic equation of state with an adiabatic index $\gamma$, and
a sufficiently high Mach number of the upstream flow, we obtain the
downstream mass density $\rho_{1}$, velocity measured in the shock frame
$u_{1}'$ and pressure $p_{1}$ in the region immediately behind the shock
front as
%
\begin{eqnarray}
\rho_{1}
&=&\frac{\gamma+1}{\gamma-1}\rho_0, \nonumber \\
u_{1}'
&=&\frac{\gamma-1}{\gamma+1}V_{\rm sh}, \nonumber \\
p_{1}
&=&\frac{2}{\gamma+1}\rho_0 V_{\rm sh}{}^2,
\label{adiabatic shock}
\end{eqnarray}
%
where $\rho_0$ and $V_{\rm sh}$ are the mass density of the upstream medium
and the shock velocity, respectively.
Hereafter, we set $\gamma=5/3$.
From the conservation equations of mass flux and momentum flux,
the mass density and pressure in the region behind the end of the cooling layer
are represented as a function of the velocity $u_2'$,
%
\begin{eqnarray}
\rho_2
&=& \frac{\rho_{1}u_{1}'}{u_2'}
= \frac{\gamma+1}{(\gamma-1)(1-\delta)}\rho_0, \nonumber \\
p_2
&=& (\rho_{1}u_{1}')(u_{1}'-u_2')+p_{1}
= \frac{2+(1-\gamma)\delta}{\gamma+1}\rho_0 V_{\rm sh}{}^2, \nonumber \\
\end{eqnarray}
%
where $\delta=1-u_2'/u_{1}'$.
Let the energy flux be
%
\begin{eqnarray}
F=u(\frac{\rho u{}^2}{2}+h), \nonumber
\end{eqnarray}
%
where $h$ is the enthalpy per unit volume.
We find that the fraction of energy flux lost via cooling is
%
\begin{eqnarray}
\varepsilon=1-\frac{F(u_2')}{F(u_{1}')}=\frac{\delta}{1+\gamma}
\left[2+(\gamma-1)\delta\right].
\end{eqnarray}
%
Following \citet{liang00}, we parameterize the downstream velocity measured in the shock frame $u_2'$ as
%
\begin{eqnarray}
u_2'=\frac{\gamma_1-1}{\gamma_1+1} V_{\rm sh}. \nonumber
\end{eqnarray}
%
Then, we obtain
%
\begin{eqnarray}
\delta
&=& 1-\frac{u_2'}{u_{1}'}=1-\frac{\gamma+1}{\gamma-1}\frac{\gamma_1-1}{\gamma_1+1},
\label{radiative jump delta} \\
\rho_2
&=& \frac{\gamma_1+1}{\gamma_1-1}\rho_0,
\label{radiative jump density} \\
p_2
&=& \frac{2}{\gamma_1+1} \rho_0 V_{\rm sh}{}^2,
\label{radiative jump pressure} \\
\varepsilon
&=& \frac{4(\gamma-\gamma_1)}{(\gamma_1+1)^2(\gamma-1)}.
\label{radiative jump}
\end{eqnarray}
%
Note that $\gamma_1$ is {\it not} an adiabatic index although
it gives the effective compression ratio as
%
\begin{eqnarray}
R_c=\frac{\rho_2}{\rho_0}=\frac{\gamma_1+1}{\gamma_1-1}.
\label{r_c}
\end{eqnarray}
%

\begin{figure}
    \center
	\includegraphics[width=10cm]{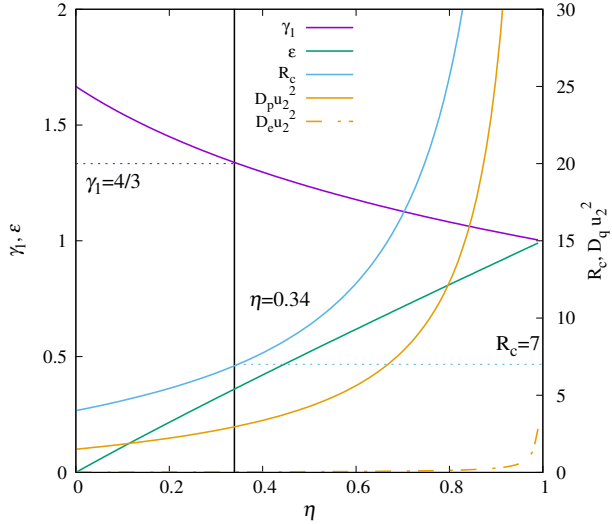}
    \vspace{15mm}
    \caption{The relationships between $\eta$ and $\gamma_1,~\varepsilon,~R_c$ and $D_q u_2{}^2=m_q u_2{}^2/(2kT_q)$.
    The left hand side vertical axis represents $\gamma_1$ and $\varepsilon$, and the right hand
    side shows $R_c$ and $D_q u_2{}^2$. The purple line is $\gamma_1$.
    The green line represents the energy loss fraction $\varepsilon$.
    The effective compression ratio $R_c$ is shown by the light blue line.
    The orange solid line is $D_{\rm p}u_2{}^2$ and the orange broken line is $D_{\rm e}u_2{}^2$ for $\beta=0.05$.
    The vertical black line in the panel is $\eta=0.34$, where $\gamma_1=4/3$ and $R_c=7$.
    }
    \label{fig:radiative}
\end{figure}

Following \citet{ghavamian02} and \citet{heng07},
we assume the downstream temperature of protons ($T_{\rm p}$)
and electrons ($T_{\rm e}$) are related to the shock velocity, $V_{\rm sh}$,
and given by
%
\begin{eqnarray}
kT_{\rm p}
&=& (1-\eta)\frac{2(\gamma-1)}{(\gamma+1)^2}
\left(
\mu_\odot f_{\rm eq}+1-f_{\rm eq}
\right)
m_{\rm p}V_{\rm sh}^2, \nonumber \\
&\equiv& (1-\eta) \frac{2(\gamma-1)}{(\gamma+1)^2}
\mu m_{\rm p}V_{\rm sh}^2,
\label{T_p} \\
kT_{\rm e}
&=& (1-\eta)\frac{2(\gamma-1)}{(\gamma+1)^2}
\left\{
\mu_\odot f_{\rm eq}+\frac{m_{\rm e}}{m_{\rm p}}\left(1-f_{\rm eq}\right)
\right\}
m_{\rm p}V_{\rm sh}^2 \nonumber \\
&\equiv& \beta kT_{\rm p}
\label{T_e}
\end{eqnarray}
%
respectively.
The definition of the energy loss rate $\eta$ is the same as Eq. \eqref{eta}.
We additionally define the temperature ratio $\beta=T_{\rm e}/T_{\rm p}$.
The $\mu_\odot=0.62$ is the mean molecular weight for solar abundances.
The situation $f_{\rm eq}=1$ ($f_{\rm eq}=0$) represents temperature equilibration (non-equilibration) among
all the particles in the fluid.
Here we consider the case in which $\alpha$ particles are in the temperature equilibrium.
Thus, for $f_{\rm eq}=1$, the mean molecular weight coincides with the value for solar abundances.
The effective mean molecular weight, $\mu\equiv\mu_\odot f_{\rm eq}+1-f_{\rm eq}$, is rewritten as a function of $\beta$,
%
\begin{eqnarray}
\mu = 1-(1-\mu_\odot)\frac{\beta-\frac{m_{\rm e}}{m_{\rm p}}}{\mu_\odot+(1-\mu_\odot)\beta-\frac{m_{\rm e}}{m_{\rm p}}}.
\end{eqnarray}
%
Equations \eqref{radiative jump
density} and \eqref{radiative jump pressure} also give the downstream proton
temperature as
%
\begin{eqnarray}
kT_{\rm p}=\mu m_{\rm p}\frac{p_2}{\rho_2}
=\frac{2(\gamma_1-1)}{(\gamma_1+1)^2} \mu m_{\rm p} V_{\rm sh}{}^2,
\end{eqnarray}
%
so that we obtain a quadratic equation for $\gamma_1$,
%
\begin{eqnarray}
& \eta
= 1-\frac{(\gamma+1)^2}{\gamma-1}\frac{\gamma_1-1}{(\gamma_1+1)^2}.
\label{quadratic}
\end{eqnarray}
%
We solve Eq. \eqref{quadratic} as
%
\begin{eqnarray}
\begin{split}
& \gamma_1
=
\frac{1}{1-\eta}
\left[
\left\{\frac{1}{2}\frac{(\gamma+1)^2}{\gamma-1}-1+\eta\right\} \right. \\
& \left. -
\sqrt{ \left\{\frac{1}{2}\frac{(\gamma+1)^2}{\gamma-1}-1+\eta\right\}^2
-(1-\eta)\left\{\frac{(\gamma+1)^2}{\gamma-1}+1-\eta\right\}}~
\right].
\end{split}
\label{gamma1}
\end{eqnarray}
%
We take the minus sign in front of the square root in Eq. \eqref{gamma1} to
derive the physical solution satisfying $\gamma_1=\gamma$ for $\eta=0$.
Hence, the compression ratio $R_c$ is given by the energy loss rate $\eta$.
The downstream velocity in the region behind the cooling layer $u_2$, which is
measured in the upstream frame, is derived from Eqs. \eqref{r_c} and \eqref{T_p} as
%
\begin{eqnarray}
u_2=\left(1-\frac{1}{R_c}\right)V_{\rm sh}
=\left(1-\frac{1}{R_c}\right)
\sqrt{\frac{(\gamma+1)^2}{2(\gamma-1)}\frac{kT_{\rm p}}{(1-\eta)\mu m_{\rm p}} }.
\label{u_2}
\end{eqnarray}
%
Figure \ref{fig:radiative} shows $\gamma_1,~\varepsilon,~R_c$ and $D_q
u_2{}^2=m_q u_2{}^2/(2kT_q)$ as function of $\eta$.
The representative value of $\eta=0.34$ is illustrated by the vertical black line, where $\gamma_1=4/3$ and $R_c=7$.
We predict that highly polarized Balmer line emissions come from large $D_q u_2{}^2$.
From the above formulae, setting the
parameters $T_{\rm p},~\eta$ and $\beta$, we calculate the polarization
degree from Eqs. \eqref{Q} and \eqref{I}.
Note that for given downstream proton temperature $T_{\rm p}$,
a large energy loss rate $\eta$ corresponds to a large shock velocity $V_{\rm sh}$. 
%
%
\par
\begin{figure}
    \center
	\includegraphics[width=10cm]{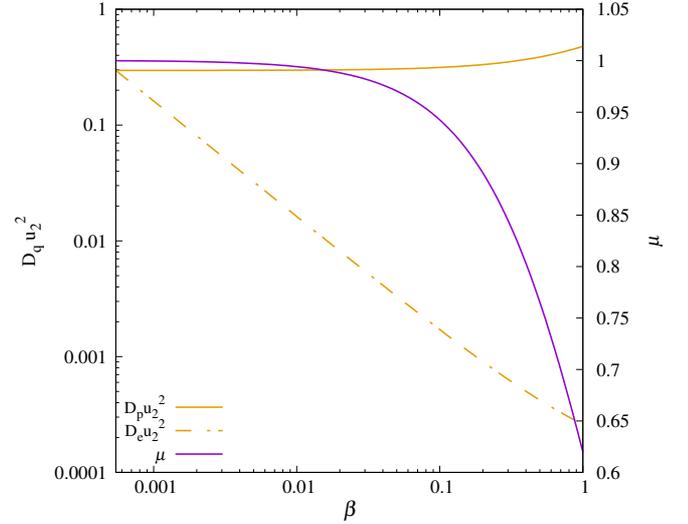}
    \vspace{15mm}
    \caption{The anisotropy of velocity distribution of protons and electrons, $D_q u_2{}^2=m_q u_2{}^2/(2kT_q)$,
    and the effective mean molecular weight $\mu$ as function of $\beta$ for $\eta=0$.
    The left hand side vertical axis represents $D_q u_2{}^2$.
    The orange solid line is $D_{\rm p}u_2{}^2$ and the orange broken line is $D_{\rm e}u_2{}^2$.
    The purple line shows $\mu$, whose value is represented by the right hand side vertical axis.
    }
    \label{fig:aniso}
\end{figure}

For typical young SNR, the temperature ratio, $\beta$, is estimated by the
intensity ratio of the broad component of H~$\alpha$ to narrow one, and to be
$\beta\sim0.03\mathchar`-0.07$ \citep[e.g.][]{adelsberg08}. Furthermore,
\citet{laming90} showed that the polarized intensity depends on the proton
temperature rather than the electron temperature. This fact arises from the
stronger anisotropy of the proton's velocity distribution than that for the
electrons, $D_{\rm e}u_2{}^2/(D_{\rm p}u_2{}^2)=m_{\rm e}/(m_{\rm_p}\beta)\ll1$.
Hence, the polarization intensity, $Q_{\rm n}$, is
mainly determined by the proton impacts. Indeed, the anisotropy of electron
velocity distribution is very small as $D_{\rm e}u_2{}^2\approx m_{\rm
e}/\left(m_{\rm p}\beta(1-\eta)\right)\ll1$. Since the electrons colliding
with energy $E_{\rm e}\ga10~{\rm eV}$ (equivalently $v_{\rm
e}\approx2500~{\rm km/s}$) excite the hydrogen atom, it contributes to
$Q_{\rm n}$ in the case of $u_2\ga2500~{\rm km/s}$ and $\beta\approx m_{\rm
e}/m_{\rm p}$.
However, the electron impacts yield unpolarized emission, that is, the
polarization degree $Q_{\rm n}/I_{\rm n}$ depends on the electron
temperature.
Figure \ref{fig:aniso} shows $D_{q}u_2{}^2$ and $\mu$ as function of $\beta$ for $\eta=0$.
%
%

\section{The Polarization degree of H~$\alpha$ Emission}
In this section, we show the results of the observed polarization degree
of H~$\alpha$ emission.
\par
%
\begin{figure}
    \center
	\includegraphics[width=10cm]{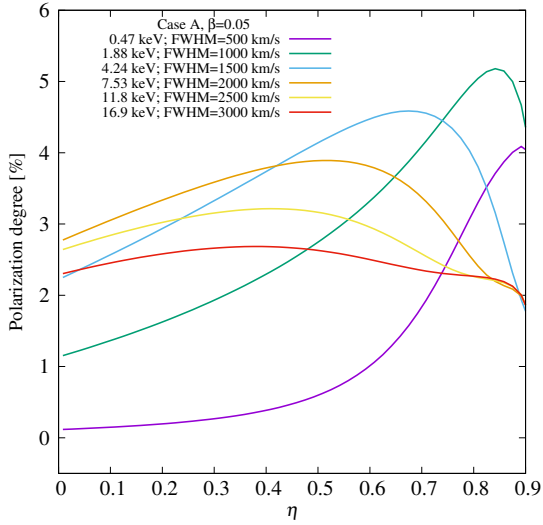}
    \vspace{15mm}
    \caption{The polarization degree of H~$\alpha$ as a function of $\eta$
    for fixed values of $T_{\rm p}$ ($0.47,~1.88,~4.24,~7.53,$ and $16.9~{\rm keV}$)
    with given $\beta=0.05$ and $\chi=\pi/2$ for Case A.
    }
    \label{fig:alpha eta}
\end{figure}
%
First of all, we show the results for $\chi=\pi/2$ and $\beta=0.05$. Figure
\ref{fig:alpha eta} represents the observed polarization degree as a function
of the energy loss rate $\eta$ for Case A with fixed $T_{\rm p}$. The solid
lines show the results for $T_{\rm p}=0.47\mathchar`-16.9~{\rm keV}$
(corresponding points are shown in the panel). For large $\eta$, the
anisotropy of the proton velocity distribution becomes large (as shown in
Fig. \ref{fig:radiative}), resulting in larger polarization degree.
For fixed $T_{\rm p}$, large $\eta$ yields large downstream velocity $u_2$ (see Eq. \eqref{u_2}).
It means that the peak of the particle velocity distribution slides to the high velocity side but its width is fixed.
When the downstream velocity $u_2$ is larger than $\approx2500~{\rm km~s^{-1}}$,
the excitation rate of the hydrogen atoms by the electron impact becomes large,
because almost all the electrons can excite the hydrogen atoms. That causes
the large unpolarized intensity $I_{\rm n}$ and the small polarization degree
$Q_{\rm n}/I_{\rm n}$.
\par
%
\begin{figure}
    \center
	\includegraphics[width=10cm]{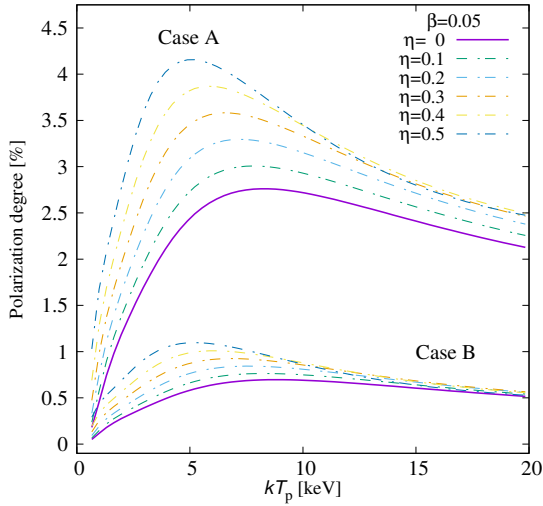}
    \vspace{15mm}
    \caption{The polarization degree of H~$\alpha$ as a function of $T_{\rm p}$
    for fixed values of $\eta$ ($0,~0.1,~0.2,~0.3,~0.4,$ and $0.5$)
    with given $\beta=0.05$ and $\chi=\pi/2$ for Case A and Case B.
    The magenta lines show the result for $\eta=0$ and the dashed lines are the results for
    $\eta=0.1\mathchar`-0.5$ from bottom to top.
    }
    \label{fig:alpha temp}
\end{figure}
%
Figure \ref{fig:alpha temp} represents the temperature dependence of the
observed polarization degree for Cases A and B. The solid lines show the
results of $\eta=0$ and the dashed lines represent $\eta=0.1\mathchar`-0.5$
from bottom to top. In Case B, the observed polarization degree is reduced
due to the Lyman line trapping, yielding larger $I_{\rm n}$.
As shown in Figures \ref{fig:alpha eta} and \ref{fig:alpha temp},
the significant energy loss rate is realized when
the observed polarization degree is $\sim4\mathchar`-5$ per cent ($\sim1$ per cent) for Case A (Case B).
\par
%
\begin{figure}
    \center
	\includegraphics[width=10cm]{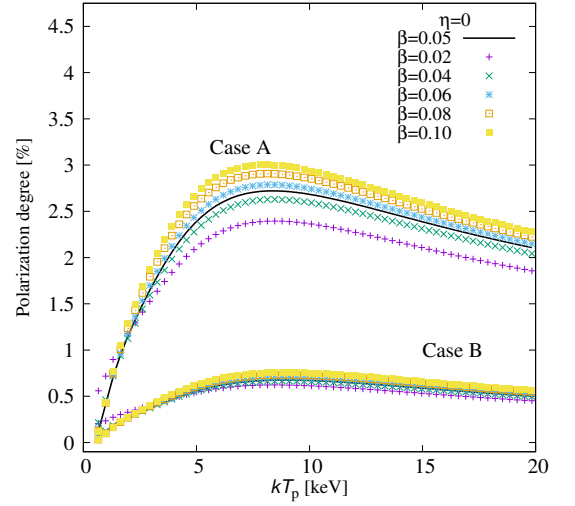}
    \vspace{15mm}
    \caption{The polarization degree of H~$\alpha$ as a function of $T_{\rm p}$
    for fixed values of $\beta$
    ($0.02,~0.04,~0.05,~0.06,~0.08$ and $0.10$)
    with given $\eta=0.05$ and $\chi=\pi/2$ for Case A and Case B.
    The black lines show the result for $\beta=0.05$ and the dashed lines are the results for
    $\beta=0.02\mathchar`-0.1$ from bottom to top.}
    \label{fig:ele temp}
\end{figure}
%
We discuss the dependence of the polarization degree on $\beta$ and $\chi$.
Figure \ref{fig:ele temp} represents the observed polarization degree for
Cases A and B for various fixed $\beta$. The solid lines correspond to the
representative value of $\beta=0.05$. The results of
$\beta=0.02\mathchar`-0.1$ are shown with points from bottom to top. For
$T_{\rm p}\ga5~{\rm keV}(u_2/1600~{\rm km/s})^2$, a large fraction of
electrons have an energy $E_{\rm e}\ga10~{\rm eV}$.
Therefore, the $\beta$ dependence is relatively large especially for Case A.
On the other hand, for Case B,
the effective cross section on electron impact $\sigma_{3p,{\rm e}}$ is dominant (see the green
and black curves in Figure \ref{fig:cross3}). Since the excitation rate is
proportional to $v_{\rm e}\sigma_{3p,{\rm e}}$, which is almost constant,
the electron temperature dependence becomes weak.
\par
%
\begin{figure}
    \center
    \includegraphics[width=10cm]{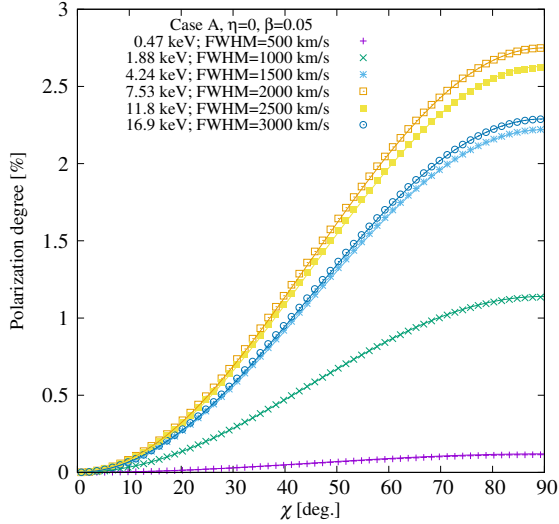}
    \vspace{15mm}
    \caption{The polarization degree of H~$\alpha$ as a function of the viewing angle $\chi$
    for fixed values of $T_{\rm p}$ ($0.47,~1.88,~4.24,~7.53,$ and $16.9~{\rm keV}$)
    with given $\beta=0.05$ and $\eta=0$ for Case A.
    The solid lines are $(Q_{\rm n}/I_{\rm n})\big|_{\chi=\frac{\pi}{2}}\sin^2\chi$.}
    \label{fig:ele view}
\end{figure}
%
The dependence on the viewing angle is shown in Figure \ref{fig:ele view}.
The points show $T_{\rm p}=0.47\mathchar`-16.9~{\rm keV}$ from bottom to top.
The solid lines are $(Q_{\rm n}/I_{\rm
n})\big|_{\chi=\frac{\pi}{2}}\times\sin^2\chi$. The unpolarized intensity
$I_{\rm n}$ is mainly determined by the electron impact due to the faster
electron velocity than proton one. Since the velocity distribution of
electron is nearly isotropic, the unpolarized intensity does not depend on
the viewing angle. Thus, the polarization degree follows $Q_{\rm n}/I_{\rm
n}\propto\sin^2\chi$ (see Eq. \eqref{Q}).
\par
\begin{figure}
    \center
    \includegraphics[width=10cm]{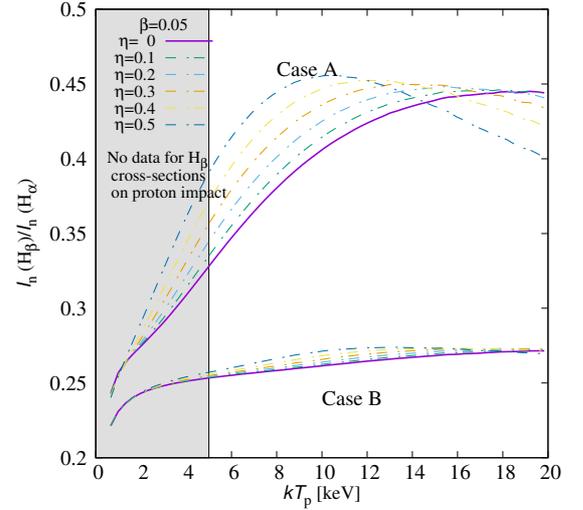}
    \vspace{15mm}
    \caption{The total intensity ratio $I_{\rm n}({\rm H_\beta})/I_{\rm n}({\rm H_\alpha})$
    as a function of $T_{\rm p}$ for fixed values of $\eta$ ($0,~0.1,~0.2,~0.3,~0.4,$ and $0.5$)
    with given $\beta=0.05$ and $\chi=\pi/2$ for Case A and Case B.
    The magenta lines show the result for $\eta=0$ and the dashed lines are the results for
    $\eta=0.1\mathchar`-0.5$ from bottom to top.
    The gray region indicates that the lack of cross section data for proton impact
    significantly affects the results, which are not reliable.
    }
    \label{fig:alpha dec}
\end{figure}
\begin{figure}
    \center
    \includegraphics[width=10cm]{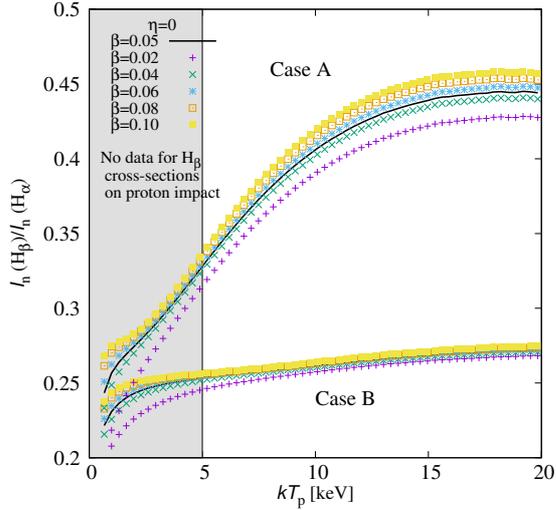}
    \vspace{15mm}
    \caption{The total intensity ratio $I_{\rm n}({\rm H_\beta})/I_{\rm n}({\rm H_\alpha})$
    as a function of $T_{\rm p}$
    for fixed values of $\beta$
    ($0.02,~0.04,~0.05,~0.06,~0.08,$ and $0.10$)
    with given $\eta=0.05$ and $\chi=\pi/2$ for Case A and Case B.
    The magenta lines show the result for $\beta=0.05$ and the dashed lines are the results for
    $\beta=0.02\mathchar`-0.1$ from bottom to top.
    The gray region indicates that the lack of cross section data for proton impact
    significantly affects the results, which are not reliable.}
    \label{fig:ele dec}
\end{figure}

Figure \ref{fig:alpha dec} shows the total intensity ratio, $I_{\rm n}({\rm
H_\beta})/I_{\rm n}({\rm H_\alpha})$ as a function of $T_{\rm p}$ for
$\chi=\pi/2$ and $\beta=0.05$ with fixed $\eta$. The dashed lines represent
$\eta=0\mathchar`-0.5$ from bottom to top. In Case A, the ratio of total
cross sections, $\sigma_{\rm tot,p}\big|_{I_{\rm n}({\rm
H_\beta})}/\sigma_{\rm tot,p}\big|_{I_{\rm n}({\rm H_\alpha})}$, is an
increasing function of temperature in the range $1000~{\rm km~s^{-1}}\la
v_{\rm p}\la 2000~{\rm km~s^{-1}}$ (equivalently, $2~{\rm keV}\la kT_{\rm
p}\la 7.5~{\rm keV}$). Therefore, the total intensity ratio is increasing
with $T_{\rm p}$. On the other hand, in Case B, $I_{\rm n}({\rm H_\alpha})$
increases by a factor of $\sim1.5$ because of $B_{3{\rm p},2{\rm s}}=1$ and
$B_{4{\rm p},3{\rm s}}\approx1$ (see the green and black curves in Figures
\ref{fig:cross3} and \ref{fig:cross4}). Therefore, the value of $I_{\rm
n}({\rm H_\beta})/I_{\rm n}({\rm H_\alpha})$ is suppressed. Moreover, the
ratio of total cross sections, $\sigma_{\rm tot,p}\big|_{I_{\rm n}({\rm
H_\beta})}/\sigma_{\rm tot,p}\big|_{I_{\rm n}({\rm H_\alpha})}$, is almost
constant for $1000~{\rm km~s^{-1}}\la v_{\rm p}\la 2000~{\rm km~s^{-1}}$.
Thus, the intensity ratio is constant with $T_{\rm p}$. Likewise, the ratio
depends on the electron temperature (Figure \ref{fig:ele dec}).
%
%
%
%
\section{The Ratio of Balmer Polarized Intensities}
The polarization degree $Q_{\rm n}/I_{\rm n}$ of Balmer line emission depends
on the effective branching ratio, $B_{nl,n'l'}$, which includes the effect of
Lyman line trapping. On the other hand, the polarized intensity $Q_{\rm n}$
is determined by the intrinsic $B_{nl,n'l'}$, which only depends on
the spontaneous transition rates. Therefore, the polarized intensity ratio of
Balmer line emission is not affected by Lyman line trapping. Moreover, the
dependence of the viewing angle is also weak (see Eq. \eqref{Q}).
In addition, the electron velocity distribution is usually isotropic in SNRs
for $\beta\ga0.01$ (see Figure \ref{fig:aniso}).
Thus, the electron temperature does not affect $Q_{\rm n}$.
Hence, the polarized intensity ratio measurements could be better than the
measurements of the polarization degree for the estimation of $\eta$.
\par
\begin{figure}
    \center
	\includegraphics[width=10cm]{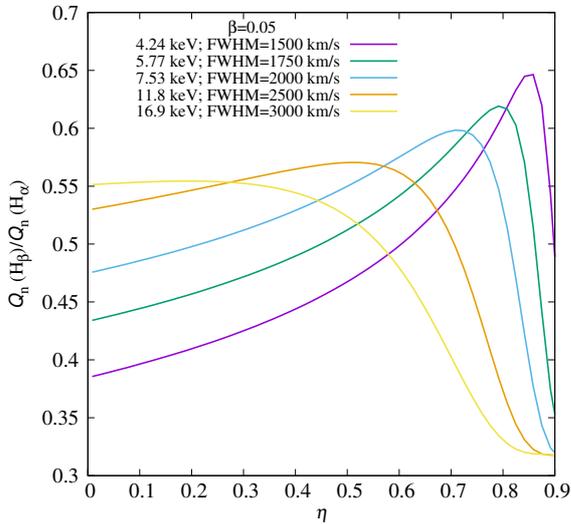}
    \vspace{15mm}
    \caption{The polarized intensity ratio $Q_{\rm n}({\rm H_\beta})/Q_{\rm n}({\rm H_\alpha})$
    as a function of $\eta$
    for fixed values of $T_{\rm p}$ ($=4.24,~5.77,~7.53,~11.8,$ and $16.9~{\rm keV}$)
    with given $\beta=0.05$.
    }
    \label{fig:beta Qeta}
\end{figure}
\begin{figure}
    \center
	\includegraphics[width=10cm]{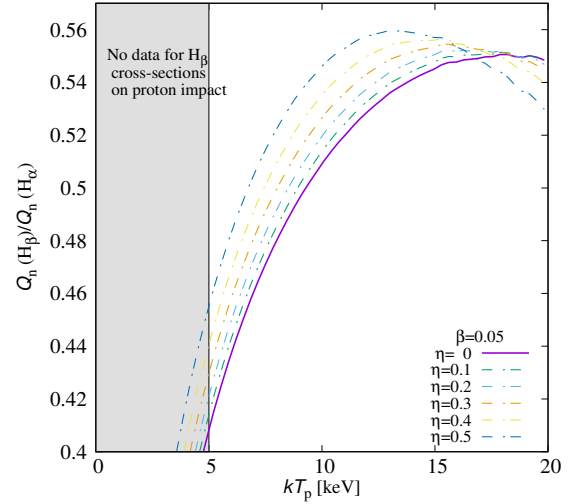}
    \vspace{15mm}
    \caption{The polarized intensity ratio
    $Q_{\rm n}({\rm H_\beta})/Q_{\rm n}({\rm H_\alpha})$
    as a function of $T_{\rm p}$
    for fixed values of $\eta$ ($0,~0.1,~0.2,~0.3,~0.4,$ and $0.5$)
    with given $\beta=0.05$.
    The magenta lines show the result for $\eta=0$ and the dashed lines are the results for
    $\eta=0.1\mathchar`-0.5$ from bottom to top.
    The gray region indicates that the lack of cross section data for proton impact
    significantly affects the results, which are not reliable.
    }
    \label{fig:beta temp}
\end{figure}
\begin{figure}
    \center
    \includegraphics[width=10cm]{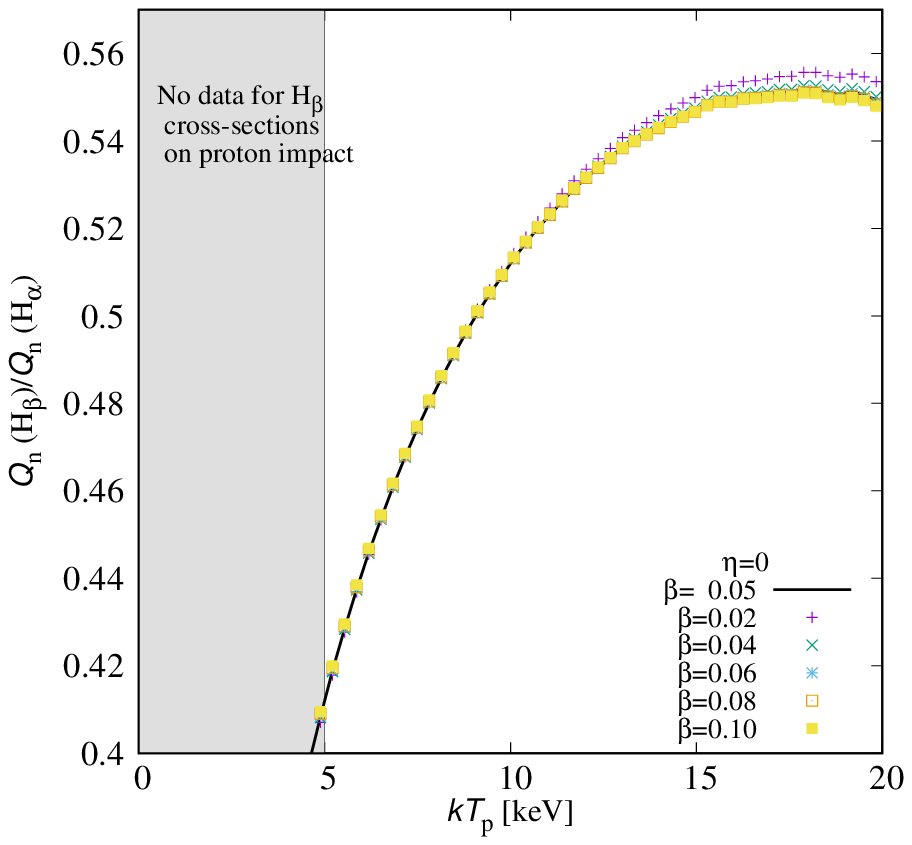}
    \vspace{15mm}
    \caption{The polarized intensity ratio
    $Q_{\rm n}({\rm H_\beta})/Q_{\rm n}({\rm H_\alpha})$
    as a function of $T_{\rm p}$
    for fixed values of $\beta$ ($0.02,~0.04,~0.05,~0.06,~0.08,$ and $0.10$)
    with given $\eta=0$.
    The black line shows the result for $\beta=0.05$ and the points are the results for
    $\beta=0.02\mathchar`-0.1$, that overlap the black line.
    The gray region indicates that the lack of cross section data for proton impact
    significantly affects the results, which are not reliable.
    }
    \label{fig:beta Qele}
\end{figure}

Figure \ref{fig:beta Qeta} shows the polarized intensity ratio of H~$\beta$
to H~$\alpha$, $Q_{\rm n}({\rm H_\beta})/Q_{\rm n}({\rm H_\alpha})$, as a
function of the energy loss rate $\eta$ for $\beta=0.05$ and $T_{\rm
p}=0.47\mathchar`-16.9~{\rm keV}$ (corresponding points are showed in the
panel). The $T_{\rm p}$ dependence of $Q_{\rm n}({\rm H_\beta})/Q_{\rm
n}({\rm H_\alpha})$ is plotted in Figure \ref{fig:beta temp}. The lines show
$\beta=0.05$ and $\eta=0\mathchar`-0.5$ from bottom to top. In particular,
the ratio is increasing with $\eta$ for $kT_{\rm p}\la15~{\rm keV}$. The
ratio of cross sections for proton impact of H~$\beta$ to H~$\alpha$,
$\sigma_{{\rm tot,p}}({\rm H_\beta})/ \sigma_{{\rm tot,p}}({\rm H_\alpha})$,
is increasing with $v_{\rm p}$ for $v_{\rm p}\la4000~{\rm km/s}$.
Since higher loss rates $\eta$ yield a larger number of high velocity protons
with fixed $T_{\rm p}$, the polarized intensity ratio is large.
Figure \ref{fig:beta Qele} shows the
ratio for different values of $\beta$, where all points with different colors
are close to with each other. Thus, the polarized intensity ratio is not
affected by $\beta$.
%
%
%
%
%
%

\section{Summary and discussion}
We have studied the linearly polarized Balmer line emission from the shocks
that efficiently accelerate CRs. Our calculation has been generalized for
arbitrary viewing angle. The Balmer line emission is polarized when
collisions between the hydrogen atoms and the charged particles
are anisotropic.
In the downstream region of the shock
with shock velocity $V_{\rm sh}$, the charged particles (in particular protons)
collide with the hydrogen atoms as a mildly-collimated beam
in the rest frame of the hydrogen atoms. When a large fraction of
SNR shock energy goes into CRs, the downstream temperature
is lower than the adiabatic case without CR acceleration, resulting in a more
anisotropic velocity distribution of charged particles and higher polarization
degree.
In other words,
for a given downstream temperature which is measured by the line width of the broad component of the H~$\alpha$ emission,
a large energy loss rate means a larger shock velocity
than the prediction of Rankine-Hugoniot
relations for adiabatic shocks, and consequently larger anisotropy of the velocity distribution.
We have found that a higher energy loss rate $\eta$, which is
defined in Eq. \eqref{T_p}, yields higher polarized Balmer line intensity. In
order to discriminate between Cases A or B in the optical depth of the Lyman
lines, the total intensity ratio so-called Balmer decrement, $I_{\rm n}({\rm
H_\alpha})/I_{\rm n}({\rm H_\beta})$, has been presented. Furthermore, we
have shown that the energy loss rate $\eta$ can be estimated by the polarized
Balmer line intensity ratio $Q_{\rm n}({\rm H_\beta})/Q_{\rm n}({\rm
H_\alpha})$ without uncertainties of the viewing angle, the electron
temperature and the Lyman line trapping.
\par
Since there are no cross section data on proton impact excitation to $n=4$ in
the range $v_{\rm p}\la1000~{\rm km~s^{-1}}$ (equivalently for downstream
temperature less than $\approx5~{\rm keV}$), our present results are
applicable for young SNRs whose downstream temperature is typically observed
as $T_{\rm p}\ga5~{\rm keV}$ \citep[for example $T_{\rm p}$ is $10~{\rm keV}$
and $6$--$7~{\rm keV}$ for SN~1006, Kepler and Tycho
respectively:][]{fesen89,ghavamian01,ghavamian02}. For older SNRs, $T_{\rm
p}$ is smaller than $5~{\rm keV}$ \citep[e.g. $T_{\rm p}\approx0.1~{\rm keV}$
for Cygnus Loop:][]{medina14}. An exception can be seen for young SNR, RCW~86
possibly showing a high energy loss rate, has $T_{\rm p}\approx2~{\rm keV}$
\citep{helder09}. Therefore, to measure the energy loss rate in RCW~86 by the
polarized Balmer-intensity ratio $Q_{\rm n}({\rm H_\beta})/Q_{\rm n}({\rm
H_\alpha})$, additional atomic data are necessary.
\par
\citet{cargill88} pointed out that electrons may be heated at SNR shocks by
plasma instabilities, such as Buneman and ion acoustic instabilities.
This electron heating would be anisotropic, directed along the shock
velocity, give rise to a different polarization signal in the Balmer lines.
Electron heating by lower hybrid waves in a shock precursor
\citep[e.g.][]{mcclements97,laming14} would be directed along the local
magnetic field leading to a different polarization direction.
However, the magnetic field can be highly disturbed by the CR-streaming instability at a gyroradius
scale of CRs in the GeV energy, $r_g\sim10^{13}~{\rm cm}(E/1~{\rm GeV})(B/1~{\rm \mu G})^{-1}$, \citep[e.g.][]{bell78}.
Since the length scale of this disturbance is much smaller than the size of the emission region ($\sim10^{16}~{\rm cm}$),
the magnetic field orientation becomes isotropic in the emission region.
Thus on average, the highly disturbed field makes net direction of electron-hydrogen atom collision isotropic on our line of sight.
As a result, the anisotropic heating of electrons directed along the magnetic field
does not yield net polarization of the observed Balmer line emissions.
Therefore, the present results can be valid when the magnetic field is
highly
disturbed at the scale smaller
than the mean free path of the atomic collision.
Besides, when the anisotropic electrons collide with other ionized species
such as Mg, Si, S and Fe, the polarized X-ray line emissions from these
species are detectable by future observation. We will study impacts of the
anisotropic heating on Balmer line polarization in a separate paper.
\par
\begin{figure}
    \center
	\includegraphics[width=10cm]{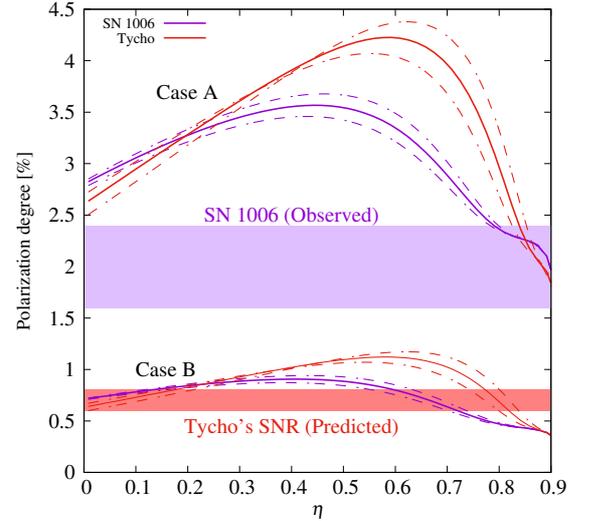}
    \vspace{15mm}
    \caption{Polarization degree of H~$\alpha$ as a function of $\eta$
    for SN 1006 (magenta) and Tycho's SNR (red).
    The broken lines indicate uncertainties of the observed proton temperature.
    The horizontal magenta belt shows the range of observed polarization degree in SN~1006, $Q_{\rm n}/I_{\rm n}=0.16\mathchar`-0.24$.
    The reddish bar illustrates predicted polarization degree in Case B for Tycho's SNR ($\eta=0.8$).
    }
    \label{fig:observe}
\end{figure}
\begin{figure}
    \center
	\includegraphics[width=10cm]{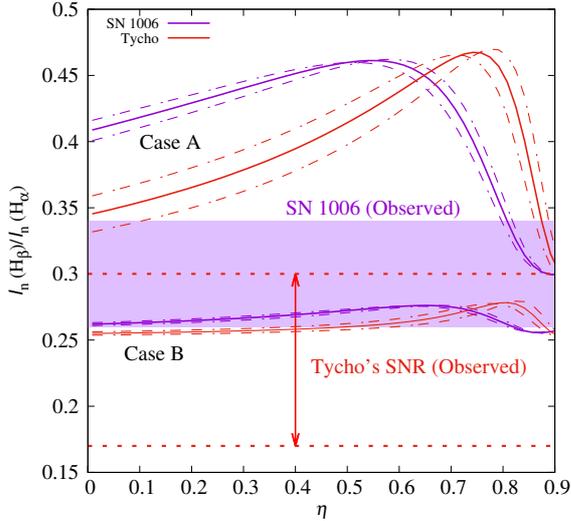}
    \vspace{15mm}
    \caption{Total intensity ratio $I_{\rm n}({\rm H_\beta})/I_{\rm n}({\rm H_\alpha})$
    for SN 1006 (magenta) and Tycho's SNR (red).
    The broken lines indicate uncertainties of the observed proton temperature.
    The observed intensity ratio in SN~1006,
    $I_{\rm n}({\rm H_\beta})/I_{\rm n}({\rm H_\alpha})=0.25\mathchar`-0.37$, is represented by horizontal magenta belt.
    The width between the two horizontal red dotted lines
    represents range of observed intensity ratio for Tycho's SNR
    ($I_{\rm n}({\rm H_\beta})/I_{\rm n}({\rm H_\alpha})=0.17\mathchar`-0.3$).}
    \label{fig:observe2}
\end{figure}
\begin{figure}
    \center
	\includegraphics[width=10cm]{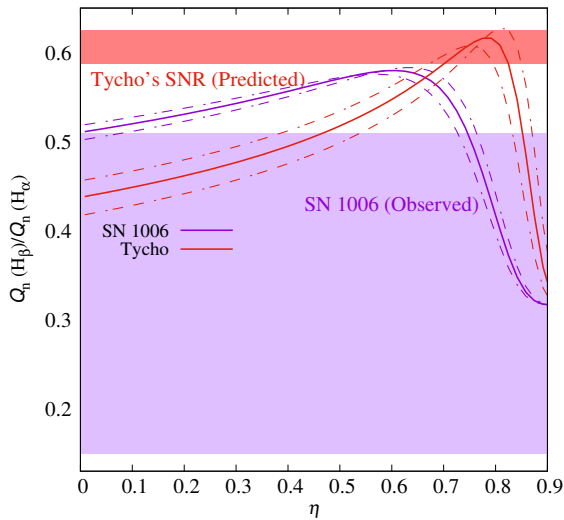}
    \vspace{15mm}
    \caption{Polarized intensity ratio $Q_{\rm n}({\rm H_\beta})/Q_{\rm n}({\rm H_\alpha})$
    for SN 1006 (magenta) and Tycho's SNR (red).
    The broken lines indicate uncertainties of the observed proton temperature.
    The reddish bar indicates predicted intensity ratio for Tycho's SNR ($\eta=0.8$).
    }
    \label{fig:observe3}
\end{figure}

For SN~1006 and Tycho's SNR, we calculate the polarization degree, the total
intensity ratio $I_{\rm n}({\rm H_\beta})/I_{\rm n}({\rm H_\alpha})$ and the
polarized intensity ratio $Q_{\rm n}({\rm H_\beta})/Q_{\rm n}({\rm
H_\alpha})$. Figure \ref{fig:observe} shows the polarization degree as a
function of $\eta$. The total (polarized) intensity ratio is represented in
Figure \ref{fig:observe2} (Figure \ref{fig:observe3}). For SN~1006 (Tycho's
SNR), we set the proton temperature $kT_{\rm p}=9.87\pm0.68$ ($kT_{\rm
p}=5.86\pm0.76$) and $\beta=0.06$ ($\beta=0.05$) following
\citet{ghavamian01,ghavamian02} and \citet{adelsberg08}. Here, the viewing
angle is fixed at $\chi=\pi/2$.
%
%
%
%
%
%
\par
\citet{sparks15} observed the polarized H~$\alpha$ emission, whose
polarization degree is $\approx2.0\pm0.4$ per cent, and that of H~${\beta}$
simultaneously in north-west region of SN~1006. If we consider Case A for
SN~1006, the observed polarization degree implies very high energy loss rate
as $\eta\ga0.8$ (see Figure \ref{fig:observe}). In this case, the total
intensity ratio $I_{\rm n}({\rm H}_\beta)/I_{\rm n}({\rm H}_\alpha)$ ranges
between $0.3$ and $0.35$ (see Figure \ref{fig:observe2}). On the other hand,
in Case B, the predicted polarization degree is smaller than $\sim1$ per cent.
As shown in Figures \ref{fig:alpha temp} and \ref{fig:observe}, the polarization
degree of H~$\alpha$ emission is significantly affected by the optical depth
of Ly~$\beta$ photon, $\tau({\rm Ly_\beta})$, which is evaluated from $I_{\rm
n}({\rm H_\beta})/I_{\rm n}({\rm H_\alpha})$. \citet{ghavamian02} analyzed
the spectra of Balmer line emissions at the same region as \citet{sparks15}
did, and obtained $I_{\rm n}({\rm H_\beta})/I_{\rm n}({\rm
H_\alpha})=0.25\mathchar`-0.37$. Combining the model of Balmer line spectra,
\citet{ghavamian02} concluded that $\tau({\rm Ly_\beta})\sim0.5$. In our
calculation for the optically thin and thick limits with $\eta=0$, the ratio
$I_{\rm n}({\rm H_\beta})/I_{\rm n}({\rm H_\alpha})$ ranges between $0.26$
and $0.41$, which is consistent with observational consequences that SN~1006
is in between Cases A and B without CR acceleration. Extending the present
model to an arbitrary $\tau({\rm Ly_\beta})$, we will precisely estimate the
energy loss rate $\eta$ from the polarization degree of H~$\alpha$ emissions,
which will be studied in the separate paper. Note that the energy loss rate
$\eta$ is also related to the ratio $Q_{\rm n}({\rm H_\beta})/Q_{\rm n}({\rm
H_\alpha})$ and independent of $\tau({\rm Ly_\beta})$. For $\eta\ga0.8$, we
predict that the polarized intensity ratio has $Q_{\rm n}({\rm
H_\beta})/Q_{\rm n}({\rm H_\alpha})=0.31\mathchar`-0.42$ (see Figure
\ref{fig:observe3}), whereas the observed value is poorly constrained,
$Q_{\rm n}({\rm H}_\beta)/Q_{\rm n}({\rm H}_\alpha)\approx
0.33\pm0.18$,
because the emission is too faint.
\par
The eastern region of Tycho's SNR has H~$\alpha$ emissions, that is known as
``knot $g$" \citep{kamper78}. The proper motion of ``knot $g$" was measured
as $\approx0.2~{\rm arcsec~yr^{-1}}$ \citep{kamper78}. On the other hand,
\cite{hayato10} observed the expansion velocity, $\approx4700~{\rm
km~s^{-1}}$, from the Doppler shift of Si X-ray line. Combining the proper
motion of the Si-rich layer ($\approx0.25~{\rm arcsec~yr^{-1}}$) measured by
\citet{katsuda10}, they concluded that the distance of Tycho's SNR is
$\approx4.0\pm1.0~{\rm kpc}$. Thus, we expect the shock velocity of the
``knot $g$" region to be $V_{\rm sh}\approx4000~{\rm km/s}$. The predicted
temperature from Rankine-Hugoniot relation, $T_{\rm RH}\approx31~{\rm keV}$,
is much higher than the observed downstream temperature, $T_{\rm
down}\approx6~{\rm keV}$. Combining these measurements, the energy loss rate
is estimated as $\eta\approx0.8$ from Eq. \eqref{eta}. Furthermore,
\citet{warren05} showed that the ratio of the forward shock radius to that of
the contact discontinuity is $1:0.93$, which implies significant energy loss
around the forward shock. This argument is independent of the uncertain
distance. However, the eastern region of Tycho's SNR (``knot $g$") was not
considered in their analysis.
\citet{ghavamian01} measured the intensity ratio $I_{\rm n}({\rm H_\beta})/I_{\rm n}({\rm H_\alpha})$ to be 0.087--0.115 (undereddened),
which becomes $0.17\mathchar`-0.3$ after correcting for the visual extinction of $1.6 \le A_v \le 3.2$,
where we take the lower and the upper limits from optical \citep{chevalier80} and X-ray absorption measurements
\citep{cassam07}, respectively.
Observational results, $\eta=0.8$ and $I_{\rm n}({\rm H_\beta})/I_{\rm n}({\rm H_\alpha})=0.17\mathchar`-0.3$,
prefer Case B.
Then, $Q_{\rm n}/I_{\rm n}\approx0.6\mathchar`-0.8$ per cent is expected.
Note that the energy
loss rate inferred from the polarization measurements does not depend on the
distance of SNR.
\par
The polarization degree depends on the electron temperature, the optical
depth and the viewing angle of the shock. These unknowns cause uncertainty of
the observed polarization degree, although the optical depth of the Lyman
line emissions and the electron temperature can be measured by observations
of $I_{\rm n}({\rm H_{\alpha}})/I_{\rm n}({\rm H_\beta})$ and the ratio of
the intensity of broad component of H~$\alpha$ emission to that of narrow
component, $I_{\rm b}({\rm H_\alpha})/I_{\rm n}({\rm H_\alpha})$. A large
energy loss rate measured by the polarization degree of H$_\alpha$ emissions
can be confirmed by the polarized intensity ratio, $Q_{\rm n}({\rm
H_\beta})/Q_{\rm n}({\rm H_\alpha})$, which does not depend on the above
values. As the energy loss rate is expected to be $\eta\approx0.8$ in Tycho's
SNR, we predict $Q_{\rm n}({\rm H_\beta})/Q_{\rm n}({\rm
H_\alpha})\approx0.60$ (see Figure \ref{fig:observe3}).
\par
\citet{shimoda15} pointed out that the density fluctuations in realistic ISM
make the SNR shock rippled and oblique everywhere. In such a situation, the
kinetic energy flux in the direction perpendicular to the shock normal is not
completely dissipated, which causes lower downstream temperature compared
with the uniform, ideal ISM case and yields apparent energy loss. The
non-dissipating kinetic energy goes to the downstream fluid motions with the
apparent energy loss rate is estimated as
$\eta\approx(\Delta\rho/\langle\rho\rangle_0)^2$ \citep[see Appendix
in][]{shimoda15}, where $\Delta\rho/\langle\rho\rangle_0$ is the amplitude of
the density fluctuation and its value is typically
$\Delta\rho/\langle\rho\rangle_0\sim0.3$ \citep[e.g.][]{inoue13}.
Therefore, the impacts of the energy loss owing to the shock rippling is
modest for the polarization degree of Balmer line emissions. The polarized
Balmer emissions from a realistic SNR shock will be studied elsewhere.
\par
We have assumed that the shock wave losing its energy owing to the CR
acceleration does not affect its shock structure, and calculated the
polarized Balmer line intensities from the downstream region. If the CRs
remain in the shock and CR pressure becomes comparable to the ram pressure in
the far upstream region, then the charged particles are decelerated by the
back reaction of CRs, leading to the modification of the shock structure in
the upstream region adjacent to the shock surface \citep[e.g.][]{berezhko99}.
The decelerated charged particles colliding the hydrogen atoms in the
upstream region are well-collimated in the rest frame of hydrogen atom (e.g.
$D_{\rm p}u_1{^2}\sim10^{2\mathchar`-3} (u_1/500~{\rm km~s^{-1}})^2(T_{\rm
p}/1\mathchar`-10~{\rm eV})^{-1}\gg1$). The Balmer line emissions from such
hydrogen atoms are also highly polarized. Therefore, detecting the polarized
Balmer line emission from the upstream region becomes evidence for the
modification of the shock structure.
We will extend the present model to study the polarized Balmer line emissions from upstream region
in forthcoming paper.

\section*{Acknowledgements}

We thank Drs. Aiko Takamine and Shota Kisaka for valuable comments to
complete this work.
We also thank the referee, John Raymond, for his comments to further improve the paper.
This work is supported by Grant-in-aids
for JSPS Fellows (15J08894, JS) and JSPS KAKENHI Grants: 16K17702 (YO),
15K05088  (RY) and 16K17673 (SK), NASA grants NNM16AA36I (Chandra GO)
and NNG16FC61I (HST GO)
and Basic research Funds of the CNR (JML).












\bsp	
\label{lastpage}
\end{document}